\documentclass[preprint]{revtex4}
\usepackage[T1]{fontenc}
\usepackage{graphicx}
\usepackage{rotating}
\usepackage{bm}        
\usepackage{amssymb}   
\usepackage{bm}
\def\etal{{\it et al.}}

\def\jcp#1#2#3{J.~Chem.~Phys.~{\bf #1},\ #2\ (#3)}
\def\cpl#1#2#3{Chem.~Phys.~Lett.~{\bf #1},\ #2\ (#3)}

\def\pra#1#2#3{Phys.~Rev.~A~{\bf #1},\ #2\ (#3)}
\def\prl#1#2#3{Phys.~Rev.~Lett.~{\bf #1},\ #2\ (#3)}

\def\jpb#1#2#3{J. Phys. B: At. Mol. Opt. Phys. {\bf #1},\ #2\ (#3)}
\def\rmp#1#2#3{Rev.~Mod.~Phys.~{\bf #1},\ #2\ (#3)}

\def\k1{k_1}
\def\k2{k_2}
\def\q1{q_1}
\def\q2{q_2}

\def\({\left (}
\def\){\right )}
\def\[{\left [}
\def\]{\right ]}

\newcommand{\beq}{\begin{equation}}
\newcommand{\eeq}{\end{equation}}

\begin{document}
\date{\today}
\flushbottom \draft
\title{
Controlling collisions of ultracold atoms with dc electric fields
}
\author{R. V. Krems\footnote{rkrems@chem.ubc.ca}}
\affiliation{
Department of Chemistry, University of British Columbia, Vancouver, B.C. V6T 1Z1, Canada
}
\begin{abstract}
It is demonstrated that elastic collisions of ultracold atoms forming a heteronuclear collision complex 
can be manipulated by laboratory practicable dc electric fields. The mechanism of electric field control 
is based on the interaction of the instantaneous dipole moment of the collision pair with external electric fields. It is shown that this interaction is dramatically enhanced in the presence of a $p$-wave shape or Feshbach scattering resonance near the collision threshold, which leads to novel electric-field-induced Feshbach resonances.

\end{abstract}

\maketitle

\clearpage
\newpage

The creation of ultracold atoms and molecules has led to many ground-breaking discoveries 
described in recent review articles \cite{reviews}. Particularly interesting is the possibility to control interactions of ultracold atoms and molecules with external electric and magnetic fields \cite{irpc}.  External field control of atomic and molecular dynamics may lead to novel spectroscopy methods, provide detailed information on mechanisms of chemical reactions and allow for the development of a scalable quantum computer \cite{demille}. 
External fields may be used to tune the scattering length for molecular interactions and alleviate the evaporative cooling of  molecules to ultracold temperatures \cite{demille_epjd}.
Collisions of ultracold atoms can be controlled by magnetic fields using Feshbach resonances \cite{feshbach} or by lasers \cite{lasers,moshe}.
Here, we show that ultracold collisions of atoms forming a heteronuclear collision complex can also be controlled by dc electric fields due to interaction of the instantaneous dipole moment of the collision pair with external fields. 
The interaction is dramatically enhanced in the presence of a $p$-wave scattering resonance near collision threshold, which allows for the possibility to tune the scattering length of ultracold atoms by laboratory 
available electric fields. This provides a new mechanism to control elastic scattering, inelastic energy transfer and chemical reactions in collisions of atoms and molecules at zero absolute temperature.

Electric field control of atomic and molecular interactions may offer several advantages over magnetic field control and optical methods. 
Optical methods to control collisions rely upon the availability of 
significant Franck-Condon factors providing couplings to electronically excited states
in a particular range of laser frequencies or a closed two-level system or the possibility to entangle 
the internal states of the colliding partners with the center-of-mass motion of the collision complex \cite{moshe}. 
Laser control of bimolecular processes may therefore be rather complicated and more system-dependent than the dc field control. 
The evaporative cooling of atoms and molecules is most easily achieved in a magnetic trap, where magnetic field control of collisions may be complicated due to varying magnetic fields of the trap. 
Electric fields induce anisotropic interactions. The study and control of anisotropic collision properties at ultracold temperatures may uncover  new phenomena in condensed matter physics - hence the recent interest in polar Bose-Einstein condensates \cite{cr}.
Anisotropic interactions may also be used to connect qubits in a quantum computer \cite{demille} and schemes for electric field control of atomic and molecular interactions are particularly relevant for 
quantum computation. 
Electric field control may be applicable to systems without magnetic moments or systems, for which $s$-wave Feshbach resonances cannot be induced in a practicable interval of magnetic fields. 
Finally, electric field control of atomic collisions may provide a sensitive probe of scattering resonances corresponding to non-zero partial waves so the mechanism described here can be used for high-precision measurements of molecular states near the dissociation threshold.

Marinescu and You \cite{marinescu} proposed to control interactions in ultracold atomic gases  by polarizing atoms with strong electric fields. The polarization changes the long-range form of the atom - atom interaction potential and modifies the scattering length. The interaction between an atom and  an electric field is, however, extremely weak and fields of as much as 250 to 700 kV/cm  were required to alter the elastic scattering cross section of ultracold atoms in the calculation of Marinescu and You. 
 The maximum dc electric field currently available in the laboratory is about 200 kV/cm \cite{meijer}.
We propose an alternative mechanism for electric field control of ultracold atom interactions and demonstrate that the scattering length of ultracold atoms can be manipulated by electric fields below 
100 kV/cm.

We consider collisions in binary mixtures of ultracold gases. Mixtures of ultracold alkali metal atoms have been created by several experimental groups \cite{stwalley,lics_experiment}.  When two different atoms collide, they form a heteronuclear collision complex which has an instantaneous dipole moment so it can interact with an external electric field. The interaction is weak. The dipole moment function of the collision complex is typically peaked around the equilibrium distance of the diatomic molecule in the vibrationally ground state and quickly decreases as the atoms separate. Only a small part of the scattering wave function samples the interatomic distances, where the dipole moment function is significant. At the same time, the interaction with an electric field couples states of different orbital angular momenta. The zero angular momentum $s$-wave motion of ultracold atoms is coupled to an excited $p$-wave scattering state, in which the colliding atoms rotate about each other with the angular momentum $l=1$ a.u. The probability density of the $p$-wave scattering wave function at small interatomic separations is very small and the coupling between the $s$- and $p$- collision states is suppressed. In this work, we show that the interaction picture is dramatically different in the presence of a $p$-wave scattering resonance near the collision threshold. The resonant enhancement of the wave function magnifies the interaction of the collision complex with an electric field so that the scattering length of the colliding atoms becomes sensitive to the electric field strength. 

To explore the effects of electric fields on atomic collisions near Feshbach resonances, we have studied the collision problem of Li with Cs in the presence of external electric and magnetic fields. Magnetic fields were used to tune Feshbach resonances.   The total Hamiltonian of two alkali metal atoms 
can be written in the form  \cite{ticknor,schmelcher}

  \begin{eqnarray}
\nonumber
 \hat{H} = - \frac{1}{2 \mu R} \frac{d^2}{dR^2} R + \frac{\hat{l}^2(\theta, \phi)}{2 \mu R^2} + 
  \gamma^{(a)} \hat{I}^{(a)} \cdot \hat{S}^{(a)} + 
  \gamma^{(b)} \hat{I}^{(b)} \cdot \hat{S}^{(b)} +
\sum_{S}  \sum_{M_S} |S M_S \rangle {V}_S(R) \langle S M_S |   - 
\\
  E  \cos{\theta} \sum_{S} \sum_{M_S} |S M_S \rangle d_S(R) \langle S M_S |  +
   2 \mu_0 B (\hat{S}_z^{(a)}    + \hat{S}_z^{(b)})
  - B \left (\frac{\mu_I^{(a)}}{I^{(a)}}  \hat{I}_z^{(a)}  + \frac{\mu_I^{(b)}}{I^{(b)}}\hat{I}_z^{(b)} \right )
  \label{ham}
  \end{eqnarray}
  
  \noindent
  where 
  $\mu$ is the reduced mass of the colliding atoms, $R$ is the interatomic separation, 
  $\theta$ and $\phi$ specify the orientation of the interatomic axis in the laboratory-fixed coordinate system,
  $l$ is the angular momentum describing the orbital motion of the colliding atoms about each other, $V_S(R)$ and $d_S(R)$ are the 
  interaction potentials and the dipole moment functions of the diatomic molecule in the states with the total electronic spin $S$, $E$ is the electric field strength, 
 $B$ is the magnitude of the magnetic field,
  $\hat{S}_z$ are the operators yielding the $z$-projection of the atomic electronic spins, $\hat{I}_z$ are the operators yielding the 
  $z$-projection of the atomic nuclear spins, $\mu_0$ is the Bohr magneton, $\mu_I$ are the nuclear magnetic moments, $\gamma$ denotes 
  the hyperfine interaction constants of the atoms, and the superscripts $(a)$ and $(b)$ are used to label the different atoms. We assume that both the magnetic 
  and electric fields are directed along the $z$-axis.

We expand the total wave function in terms of products of the partial waves $\phi_{l m_l}$ and the atomic wave functions as follows

  \begin{eqnarray}
  \psi = \frac{1}{R} \sum_{l} \sum_{m_l}  \sum_{f_a} \sum_{m_{f_a}}   \sum_{f_b} \sum_{m_{f_b}} 
  F_{l m_l f_a m_{f_a}  f_b m_{f_b}}(R) \phi_{l m_l}(\theta, \phi) \chi_{f_a m_{f_a}} \chi_{f_b m_{f_b}},
 \label{expansion}
 \end{eqnarray} 

\noindent
where $f_a$ and $f_b$ are the total angular momenta of the atoms $(a)$ and $(b)$ and $m_{f_a}$ and $m_{f_b}$ are the projections 
of $f_a$ and $f_b$ on the space-fixed quantization axis. The substitution of the expansion (\ref{expansion}) in the Schr\"{o}dinger equation 
with the Hamiltonian (\ref{ham}) leads to a system of coupled differential equations for the expansion coefficients $F(R)$. 
The coupling matrices are determined in the $|f_a m_{f_a} f_b m_{f_b} \rangle$ representation using the angular momentum 
recoupling coefficients, as described, for example, in Ref. \cite{kluwer}. We note that $f_a$ and $f_b$ are not conserved at $R=\infty$
due to interactions of the atoms with magnetic fields and the coupled equations cannot be solved directly in the 
$|f_a m_{f_a} f_b m_{f_b} \rangle$ representation. Therefore, we introduce an additional diagonalizing transformation as was 
described previously \cite{romanJCP} and solve the coupled equations in the representation, in which the Hamiltonian (\ref{ham})
is diagonal at $R=\infty$. 
For $V_S(R)$, we use the  interatomic interaction potentials of the LiCs molecule in the electronic states $^1\Sigma^+$ and $^3\Sigma^+$ constructed by analytical 
fits to the ab initio data of Korek \etal~\cite{lics} at interatomic distances between $5.2$ and $14$ bohr smoothly  joined to the long-range form $-C_6/R^6 - C_{10}/R^{10} - C_{12}/{R^{12}}$. The coefficients $C_6, C_{10}$ and $C_{12}$ for the Li-Cs interaction were taken from the database of Marinescu and Vrinceanu \cite{daniel}. For $d_S(R)$, we use the Gaussian functions that resemble the dipole moment functions of the LiCs molecule in the $S=0$ and $S=1$ states computed by Aymar and Dulieu \cite{olivier}.  The Li-Cs system is a representative example of a heteronuclear collision complex with a substantial dipole moment. Ultracold mixtures of Li and Cs have been created and studied in several recent experiments of Mudrich \etal~\cite{lics_experiment}. We consider collisions of Li and Cs atoms initially in the $m_{f_a} = m_{f_b} = +1$ states of the lowest energy. Three partial waves $l=0$, 
$l=1$ and $l=2$ were included in these calculations, which resulted in systems of 45 coupled differential equations. We have verified that adding higher partial waves does not change the results presented. The interaction with asymptotically closed channels corresponding to the quantum numbers $m_{f_a} + m_{f_b} = 2$ gives rise to Feshbach scattering resonances.

Figure 1 shows the variation of the cross sections 
for elastic $s$-wave scattering, elastic $p$-wave scattering and the $s \rightarrow p$ transition as a function of the magnetic field 
at a collision energy of $10^{-7}$ cm$^{-1}$.
The elastic cross sections were computed at zero electric field strength and the cross section 
for the $s \rightarrow p$ transition was computed at an electric field magnitude of 30 kV/cm. 
 The $s \rightarrow p$ transition is induced by the interaction of the collision complex with the electric field. 
 The peaks in the elastic $s$-wave cross section correspond to $s$-wave Feshbach resonances; the peaks in the elastic $p$-wave cross section correspond to $p$-wave Feshbach resonances. 
 The $s \rightarrow p$ cross section is enhanced at both the $s$-wave and $p$-wave resonances. This indicates that the $s \rightarrow p$ coupling and the interaction with electric fields can be enhanced either by an $s$-wave or by a $p$-wave scattering resonance near threshold. The resonance enhancement of the $s \rightarrow p$ coupling changes the cross section for the $s \rightarrow p$ transition by 4 to 6 orders of magnitude.

Figure 2 demonstrates the effect of a $p$-wave resonance on the $s$-wave elastic scattering cross section. The elastic scattering cross section is proportional to the square of the scattering length.  At zero electric field, the $s$- and $p$- states are uncoupled and the $p$-wave resonance cannot affect the scattering of ultracold atoms. As the field strength increases, the interaction with the field induces a resonant variation of the elastic $s$-wave scattering cross section. The variation is sensitive to the magnitude of the electric field. The left panels of Fig. 2 show that increasing the electric field strength from 90 to 105 kV/cm changes the cross sections for $s$-wave 
elastic  scattering and the $s \rightarrow p$ transition by many orders of magnitude. 

Feshbach resonances may not be tunable in some molecules of interest for chemical applications. However, most atomic and molecular systems are characterized by shape (single-channel) resonances. Shape resonances are ubiquitous in heavy systems which posses large densities of states. To explore the effects of electric fields on atomic collisions near shape resonances, we repeated the calculations for Li - Cs collisions using a single potential curve $^1\Sigma$ and the corresponding dipole moment function. To induce a single-channel $p$-wave resonance, we slightly modified the reduced mass (or equivalently the interaction potential) of the colliding atoms.  
Figure 3 shows the variation of the elastic scattering cross section with the electric field strength near the shape $p$-wave resonance. The two curves correspond to different positions of the
shape resonance.  
This model result demonstrates that electric fields may also be used to manipulate collisions in the presence of naturally occurring single-channels resonances.

In summary, we have shown that the scattering length of ultracold atoms forming a heteronuclear collision complex can be manipulated by laboratory moderate dc electric fields. 
The elastic scattering cross section is modified due to the interaction of the instantaneous dipole moment of the collision system with external electric fields. This interaction is dramatically enhanced in the presence of an $s$-wave or $p$-wave scattering (shape or Feshbach) resonance near the collision threshold. 

 Scattering resonances are ubiquitous in collisions of heavy atoms and molecules and many collision systems will naturally have a resonance state near collision thresholds \cite{bohn_resonances}. One example is the F + H$_2$ collision complex \cite{irpc,bodo}. In addition, both $s$-wave and $p$-wave resonances 
 can be tuned by magnetic fields. Atomic and molecular collisions 
 may thus be controlled by superimposed magnetic and electric fields. Magnetic fields do not couple different partial wave states. Tuning a $p$-wave resonance in a collision system of distinct atoms will not  affect ultracold collisions in the $s$-wave limit. Applying an electric field as discussed in this paper will then provide a route to control ultracold collisions. The scattering amplitude in the $p$-wave collision channel is not spherically symmetric so the mechanism described here can be used to manipulate angular distributions of scattering products at ultracold temperatures. 
Ultracold collisions are dominated by $s$-wave scattering and resonances corresponding to non-zero partial waves may be difficult to observe in an ultracold scattering experiment. Our results show that measurements of the $s$-wave scattering length in superimposed magnetic and electric fields may yield accurate information on molecular states with non-zero angular momenta.
 Our work suggests a mechanism for electric field control of  chemical reactions at zero temperature. Increasing the scattering length for atom - molecule collisions results in enhancement of chemical reaction rates \cite{irpc,bodo}. Consider, for example, the chemical reaction between a fluorine atom and an H$_2$ molecule \cite{bala}. The F + H$_2$ reactive complex must have a large dipole moment so external electric fields may enhance or reduce the reaction time, 
similarly to how the elastic scattering of Li with Cs has been shown to vary with the electric field 
strength. An experimental demonstration of electric field effects on ultracold chemical reactions will open up a new field of controlled chemistry with many interesting applications~\cite{irpc}.

The results of Fig. 2 show that dc electric fields may induce novel Feshbach resonances. The resonant enhancement in the left panel of Fig. 2 is due to coupling between two asymptotically degenerate states with finite lifetimes. Dimers near an electric-field resonance will be polarized and may interact through the long-range dipolar interaction. 
An experimental study of electric-field-induced resonances in ultracold gases may uncover new interesting phenomena. For example, the properties of Bose-Einstein condensates may be modified by such resonances so that the ultracold mixtures become unstable or vise versa.  In addition, the coupling between the $s$- and $p$-wave states may alter the expansion properties of ultracold clouds released from the trap - the phenomenon recently observed in the condensate of Cr \cite{cr}.

\begin{figure}[ht]
\caption{
Cross sections for elastic $s$-wave scattering (upper panel), elastic $p$-wave scattering 
(middle panel) and inelastic $s \rightarrow p$ transitions at an electric field strength 
of $30$ kV/cm. The elastic scattering cross sections are computed at zero electric field. 
The collision energy is $10^{-7}$ cm$^{-1}$. 
}
\vspace{0.5cm}
\label{fig:1}
\begin{center}
\includegraphics[scale=0.4]{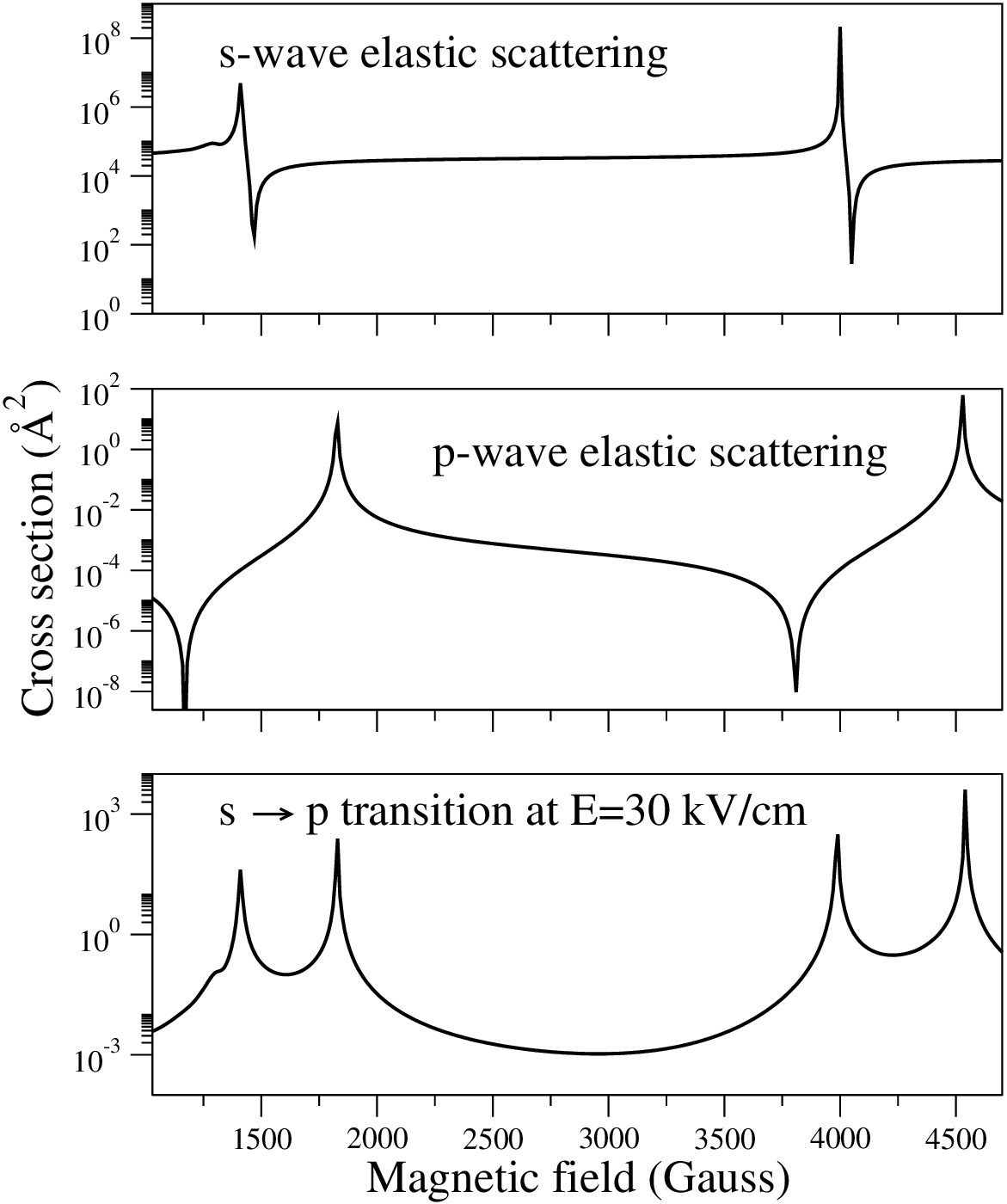}
\end{center}
\end{figure}

\begin{figure}[ht]
\caption{
Left panels: Cross sections for elastic $s$-wave scattering (upper panel),  and inelastic $s \rightarrow p$ transitions (lower panel) as a function of the electric field strength 
at the magnetic field magnitude 4709.6 Gauss. 
Right panels: Elastic $s$-wave (full curve) and $p$-wave (broken curve) cross sections
computed at zero electric field (upper panel) and 100 kV/cm (lower panel).
The collision energy is $10^{-7}$ cm$^{-1}$. 
The electric field not only couples the $s$- and $p$-states, but also shifts the positions of the $p$-wave resonances, thereby inducing the resonant variation of the $s$-wave cross section.  
}
\vspace{0.5cm}
\label{fig:2}
\begin{center}
\includegraphics[scale=0.4]{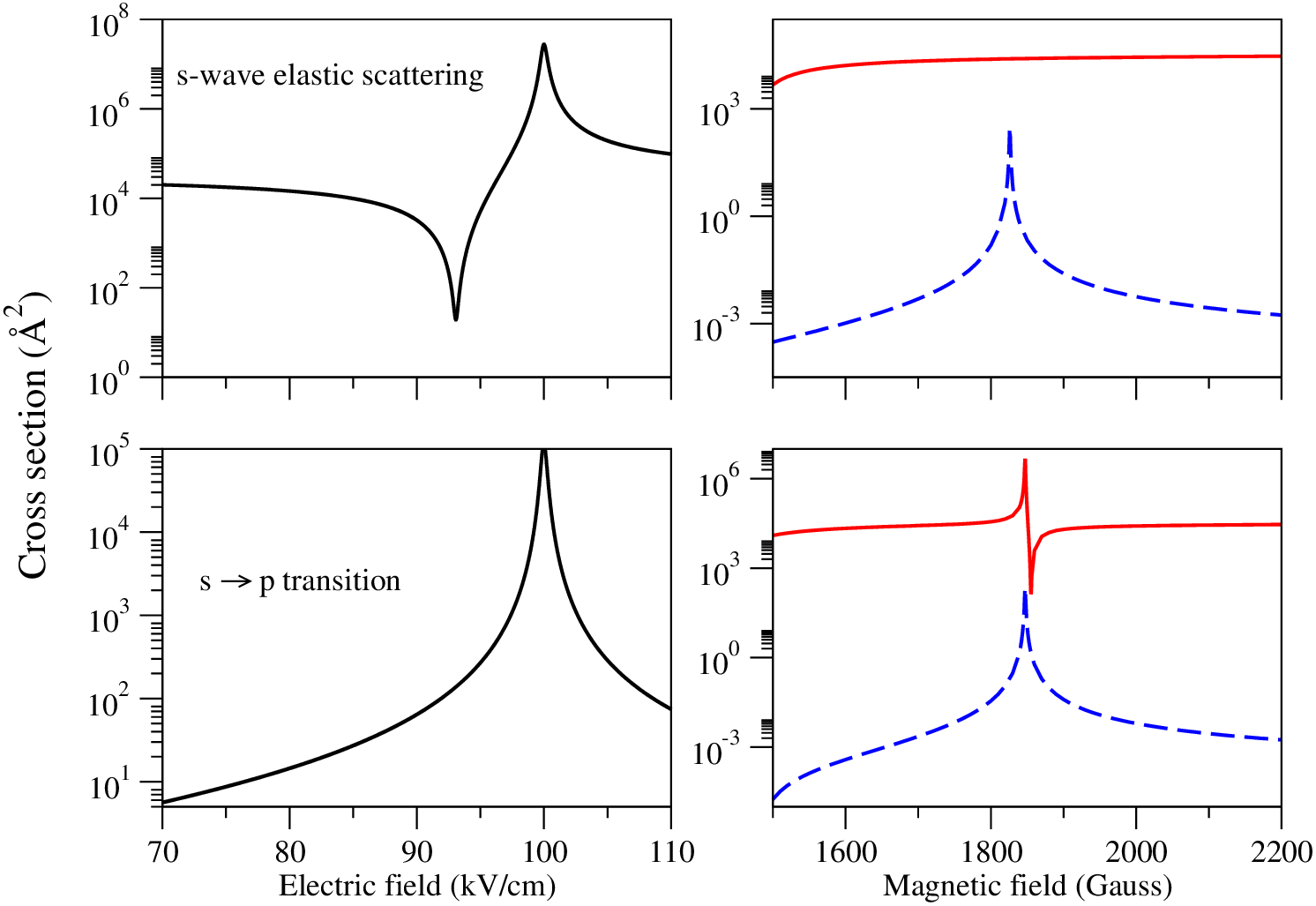}
\end{center}
\end{figure}

\begin{figure}[ht]
\caption{
Variation of elastic $s$-wave scattering cross section at zero collision energy 
with the electric field in the presence of a $p$-wave shape resonance 
near threshold. The collision energy is $10^{-7}$ cm$^{-1}$. 
The two curves correspond to two slightly different positions of the resonance. 
} 
\vspace{0.5cm}
\label{fig:3}
\begin{center}
\includegraphics[scale=0.4]{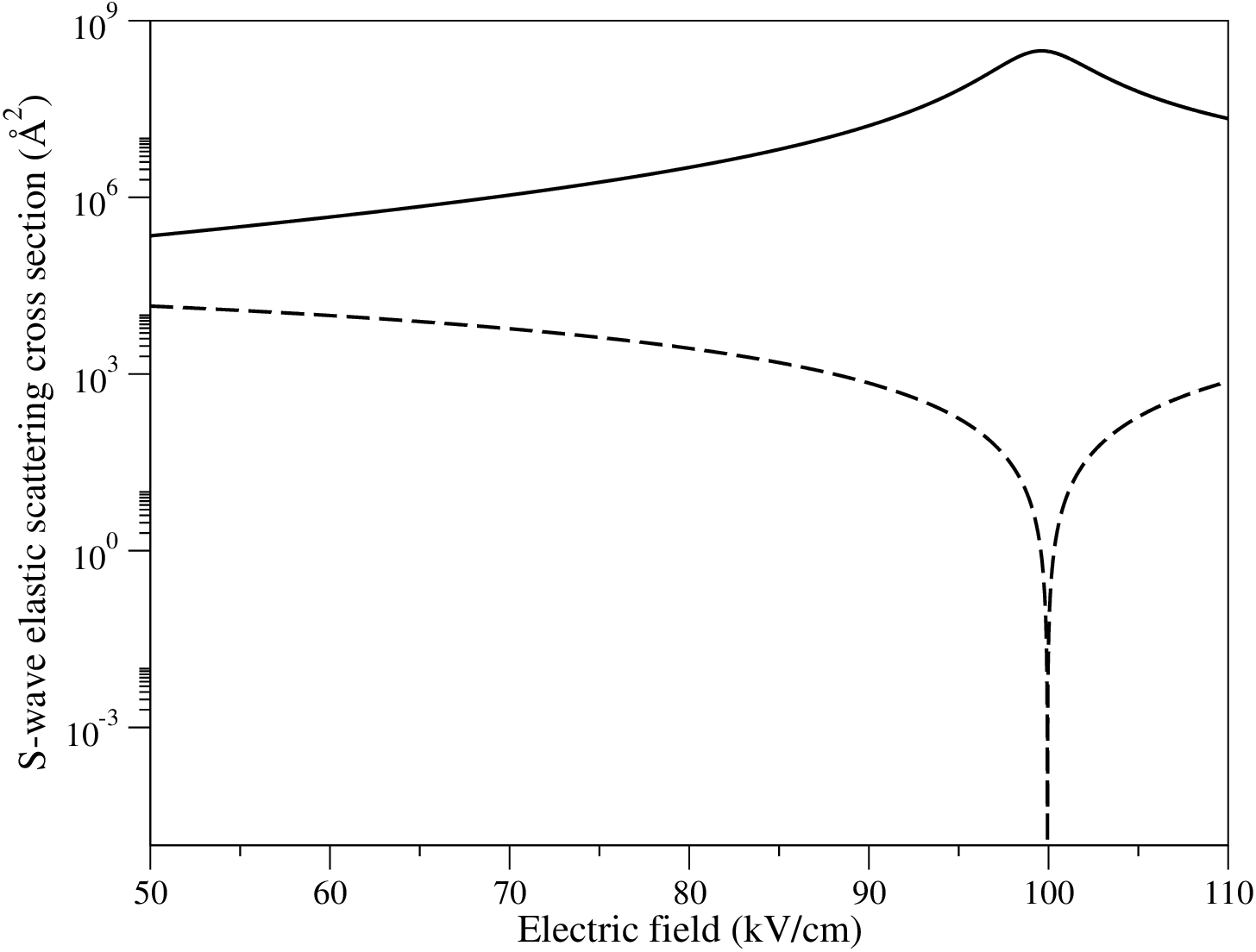}
\end{center}
\end{figure}


\clearpage
\newpage

\begin{thebibliography}{99}

\bibitem{reviews}
P.S. Julienne, Nature {\bf 424}, 24 (2003); 
J. Anglin and W. Ketterle, Nature {\bf 416}, 211 (2002);
W. Ketterle, 
 \rmp{74}{1131}{2002};
B. G. Levi, Phys. Today {\bf 53}, 46 (2000);
H.L. Bethlem and G. Meijer,  
 Int. Rev. Phys. Chem. {\bf 22}, 73 (2003);
J. Doyle, B. Friedrich, R.V. Krems, and F. Masnou-Seeuws, 
Eur. Phys. J. D {\bf 31}, 149 (2004).

\bibitem{irpc}
R.V. Krems, Int. Rev. Phys. Chem. {\bf 24}, 99 (2005). 

\bibitem{demille}
D. DeMille, \prl{88}{067901}{2002}

\bibitem{demille_epjd}
D. DeMille, D. R. Glenn, and J. Petricka, Eur. Phys. J. D {\bf 31}, 375 (2004). 


\bibitem{feshbach}
E. Tiesinga, B. J. Verhaar, and H. T. C. Stoof
Phys. Rev. A 47, 4114-4122 (1993);
 J. Stenger, S. Inouye, M. R. Andrews, H. J. Miesner, D. M. Stamper-Kurn, and 
W. Ketterle, \prl{82}{2422}{1999};
 M. Greiner, C. A. Regal, and D. S. Jin, 
Nature (London) {\bf 426}, 537 (2004); 
S. Jochim \etal, Science {\bf 302}, 2101 (2003); 
M. W. Zwierlein \etal, \prl{91}{250401}{2003};
V. Vuletic, C. Chin, A. J. Kerman, and S. Chu
\prl{83}{943}{1999}; J. L. Roberts, N. R. Claussen, S. L. Cornish, and C. E. Wieman
Phys. Rev. Lett. 85, 728-731 (2000). 





\bibitem{lasers}
P. Fedichev, Yu Kagan, G. V. Shlyapnikov, and J. T. M. Walraven, 
\prl{77}{2913}{1996};
D. J. Heinzen, R. Wynar, P. D. Drummond, and K. V. Kheruntsyan
\prl{84}{5029}{2000};
O. Mandel, M. Greiner, A. Widera, T. Rom, T. W. H\"{a}nsch, and I. Bloch,
Nature (London) {\bf 25}, 937 (2003);
 M. Theis, G. Thalhammer, K. Winkler, M. Hellwig, G. Ruff, R. Grimm, and J. Hecker Denschlag, 
\prl{93}{123001}{2004}; 


\bibitem{moshe}
M. Shapiro, and P. Brumer,  2003,  ``{\it Principles of Quantum Control of Molecular Processes}" (John Wiley and Sons, Inc., New Jersey). 






\bibitem{cr}
J. Stuhler, A. Griesmaier, T. Koch, M. Fattori, T. Pfau, S. Giovanazzi, P. Pedri, and L. Santos
Phys. Rev. Lett. {\bf 95}, 150406 (2005). 


\bibitem{marinescu}
M. Marinescu and L. You, 
\prl{81}{4596}{1998}.


\bibitem{meijer}
H. L. Bethlem, G. Berden, F. M. H. Crompvoets,  R. T. Jongma, A. J. A. van Roij , and G. Meijer, 
Nature {\bf 406}, 491 (200). 


\bibitem{stwalley}
D. Wang, J. Qi, M.F. Stone, O. Nikolayeva, B. Hattaway, S.D. Gensemer, H. Wang, W.T. Zemke, P.L. Gould, E.E. Eyler, and W.C. Stwalley, 
Eur. Phys. J. D {\bf 31}, 165 (2004); 
M. Mudrich,  O. B\"{u}nermann, F. Stienkemeier, O. Dulieu, and M. Weidem\"{u}ller, 
Eur. Phys. J. D {\bf 31}, 291 (2004);  
A. J. Kerman, J. M. Sage, S. Sainis, T. Bergeman, and D. DeMille
Phys. Rev. Lett. {\bf 92}, 033004 (2004).



\bibitem{lics_experiment}
M. Mudrich, S. Kraft, K. Singer, R. Grimm, A. Mosk, and M. Weidem\"{u}ller
Phys. Rev. Lett. {\bf 88}, 253001 (2002).
M. Mudrich, S. Kraft, J. Lange, A. Mosk, M. Weidem\"{u}ller, and E. Tiesinga
Phys. Rev. A {\bf 70}, 062712 (2004).


\bibitem{ticknor}
C. Ticknor and J. L. Bohn, arXiv: physics/0506104

\bibitem{schmelcher}
R. Gonzalez-Ferez and P. Schmelcher, 
\pra{69}{023402}{2004};
\pra{71}{033416}{2005};
Europhys. Lett. {\bf 72}, 555 (2005).

\bibitem{kluwer}
R.V. Krems and A. Dalgarno, 
  in Fundamental World of Quantum Chemistry, chapter 14, Vol. 3,  273, eds: E. Kryachko and E. Brandas (Kluwer 2004). 

\bibitem{romanJCP}
R.V. Krems and A. Dalgarno, \jcp{120}{2296}{2004}.


\bibitem{lics}
M. Korek \etal, 
Can. J. Phys. {\bf 78}, 977 (2000).

\bibitem{daniel}
D. Vrinceanu and M. Marinescu, http://cfa-www.harvard.edu/$\sim$dvrinceanu/Mircea/AlkaliMetal.html

\bibitem{olivier}
M. Aymar and O. Dulieu, 
\jcp{122}{204302}{2005}.



\bibitem{bohn_resonances}
J. L. Bohn, A. V. Avdeenkov, and M. P. Deskevich, 
\prl{89}{203202}{2002}.


\bibitem{bodo}
E. Bodo \etal, 
\jpb{37}{3641}{2004}.




\bibitem{bala}
N. Balakrishnan and A. Dalgarno,   
\cpl{341}{652}{2001}.
 


\end{thebibliography}
\end{document}